\documentclass[%
reprint,
superscriptaddress,
 amsmath,amssymb,
 aps,
pra,
]{revtex4-1}
\usepackage{graphicx}
\usepackage{dcolumn}
\usepackage{bm}

\begin{document}

\title{Sawtooth wave adiabatic passage slowing of dysprosium}

\author{N. Petersen}
\affiliation{%
 QUANTUM, Institut f\"ur Physik, Johannes Gutenberg-Universit\"at, 55099 Mainz, Germany
}
\affiliation{%
Graduate School Materials Science in Mainz, Staudingerweg 9, 55128 Mainz, Germany}

\author{F. M\"uhlbauer}%
\affiliation{%
 QUANTUM, Institut f\"ur Physik, Johannes Gutenberg-Universit\"at, 55099 Mainz, Germany
}

\author{L. Bougas}
\affiliation{Helmholtz Institut Mainz, Johannes Gutenberg-Universit\"at, Staudingerweg 18, 55128 Mainz, Germany 
}
\author{A. Sharma}
\affiliation{Helmholtz Institut Mainz, Johannes Gutenberg-Universit\"at, Staudingerweg 18, 55128 Mainz, Germany
}
\author{D. Budker}
\affiliation{%
 QUANTUM, Institut f\"ur Physik, Johannes Gutenberg-Universit\"at, 55099 Mainz, Germany
}
\affiliation{Helmholtz Institut Mainz, Johannes Gutenberg-Universit\"at, Staudingerweg 18, 55128 Mainz, Germany 
}
\affiliation{Department of Physics, University of California, Berkeley, CA 94720-7300, USA 
}
\author{P. Windpassinger}
\affiliation{%
 QUANTUM, Institut f\"ur Physik, Johannes Gutenberg-Universit\"at, 55099 Mainz, Germany
}
\affiliation{%
Graduate School Materials Science in Mainz, Staudingerweg 9, 55128 Mainz, Germany}

\begin{abstract}
We report on sawtooth wave adiabatic passage (SWAP) slowing of bosonic and fermionic dysprosium isotopes by using a 136\,kHz wide transition at 626\,nm.
A beam of precooled atoms is further decelerated in one dimension by the SWAP force and the amount of atoms at near zero velocity is measured. We demonstrate that the SWAP slowing can be twice as fast as in a conventional optical molasses operated on the same transition. In addition, we investigate the parameter range for which the SWAP force is efficiently usable in our set-up, and relate the results to the adiabaticity condition. Furthermore, we add losses to the hyperfine ground-state population of fermionic dysprosium during deceleration and observe more robust slowing with SWAP compared to slowing with the radiation pressure force.

\end{abstract}


\maketitle

Laser cooling is an important prerequisite in many research areas like quantum gases \cite{metcalf1999laser,ketterle1999making,ketterle2008making}, ion traps \cite{eschner2003laser} and optical atomic clocks \cite{ludlow2015optical}. A commonly applied method to cool atoms is radiation pressure (RAD) cooling, which is based on directed absorption of photons by the atoms and subsequent spontaneous emission. That way, temperatures in the $\mathrm{mK}$ to $\mu \mathrm{K}$ range are typically reached. The minimal temperature and the maximum force and velocity-capture range are limited by the decay rate $\Gamma$ from the upper state of the used transition \cite{metcalf1999laser}. Relying on spontaneous emission for cooling requires the transition to be closed or at least nearly closed such that repumping from ``dark states'' is feasible.

Extending laser cooling to multi-frequency light fields can overcome some of the limitations of RAD cooling \cite{metcalf2017colloquium}, particularly the maximum achievable force and velocity-capture range. Examples are the bichromatic force \cite{voitsekhovich1989observation,soding1997short,partlow2004bichromatic,yatsenko2004dressed,corder2015laser} and forces originating from pulsed rapid adiabatic passages \cite{lu2005bloch,miao2007strong,stack2011numerical}.

\begin{figure}[ht!]
	\centering
		\includegraphics{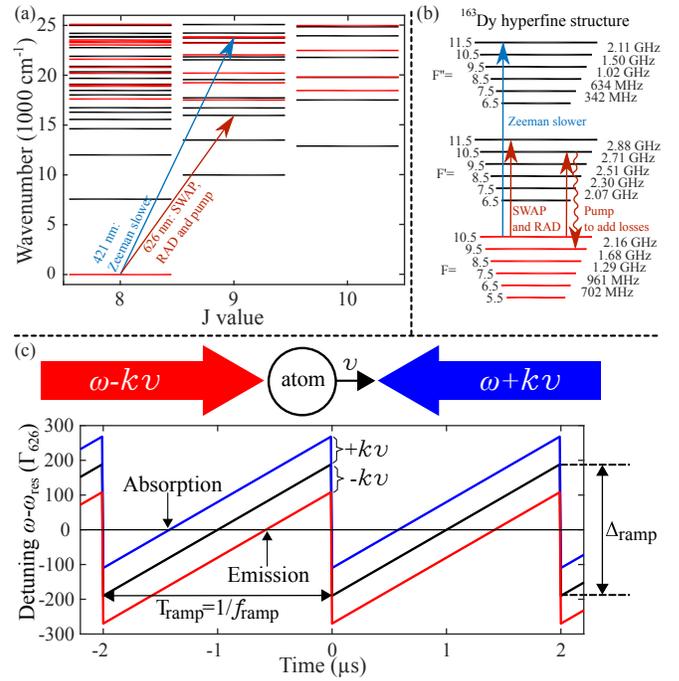}
		\caption{(Color online) (a) Excerpt of Dy energy levels for total angular momenta $\mathrm{J}=8,9,10$ \cite{NIST_ASD}. Even (odd) parity levels are drawn in red (black). (b) Hyperfine structure of the $^{163}\mathrm{Dy}$ isotope \cite{youn2010dysprosium,hogervorst1978isotope}. The arrows indicate the transitions used for the Zeeman slower at $421\,\mathrm{nm}$, for the SWAP deceleration at $626\,\mathrm{nm}$ and for the implementation of ground-state losses. (c) Schematic of the SWAP force mechanism. The black sawtooth function describes the time dependent detuning of the counter-propagating laser beams with respect to an atom at rest. An atom moving with finite velocity $v$ experiences Doppler shifted frequency ramps indicated by the blue and red lines in the plot. Important parameters of the sawtooth ramps used for deceleration of dysprosium like the ramp amplitude $\Delta_\mathrm{ramp}$ and ramp repetition rate $f_\mathrm{ramp}$ are indicated.}
	\label{fig:figure1a}
\end{figure}

Recently, a novel cooling technique called sawtooth wave adiabatic passage (SWAP) was demonstrated for strontium \cite{norcia2018narrow,muniz2018robust} on a $2 \pi \times 7.5\,\mathrm{kHz}$ wide transition and rubidium on a Raman transition \cite{greve2018laser}. In this work, we demonstrate its application for dysprosium (Dy). We use the $\Gamma_{626}=2 \pi \times 136\,\mathrm{kHz}$ transition from the ground-state at $626\,\mathrm{nm}$ (Fig.~\ref{fig:figure1a}(a)) with a saturation intensity of $\mathrm{I}_{\mathrm{S}}=72 \,\mu\mathrm{W}/\mathrm{cm}^2$ \cite{martin1978,hogervorst1978isotope,gustavsson1979lifetime} to generate the SWAP force. Thereby, we validate that the SWAP force also works for more than one order of magnitude broader transitions than previously demonstrated. 
The $626\,\mathrm{nm}$ transition is comparable in terms of linewidth and saturation intensity to the $2 \pi \times 160\,\mathrm{kHz}$ $X^2 \Sigma \rightarrow A'^2 \Delta_{3/2}$ transition in YO \cite{collopy2015prospects}. To show the robustness of the process, we exploit the hyperfine structure of fermionic $^{163}\text{Dy}$ (Fig.~\ref{fig:figure1a}(b)) and use it to induce ground-state losses during the deceleration.

Similar to the bichromatic force and pulsed rapid adiabatic passage, the SWAP force does not rely on spontaneous photon scattering to remove kinetic energy, but uses rapid adiabatic passages for the excitation as well as for the emission processes and is experimentally straightforward to implement. 
It is expected to have several advantages over the RAD force
\begin{equation}
F_{\mathrm{RAD,max}}=\frac{\hbar k}{2}\Gamma \, \mathrm{,}
\label{eq: RAD force}
\end{equation}
which is given here for a saturated transition and a resonant laser beam \cite{metcalf1999laser}. First, the maximum SWAP force is not limited by the decay rate from the upper state and, thus, in principle, even with narrow transitions strong forces can be generated when high ramp repetition rates are applied and the adiabaticity condition is fulfilled. Furthermore, SWAP cooling is expected to remove considerably more atomic momentum per spontaneously scattered photon compared to RAD cooling and hence the requirement to find a closed cooling transition could be relaxed \cite{bartolotta2018laser}. This is of special interest for cooling of molecules and in general of systems with open transitions. 
Due to fewer spontaneous decays being involved, the SWAP force may be also advantageous for cooling optically dense samples, where spontaneously emitted photons could be radiationally trapped and may reduce the cooling efficiency otherwise.

To briefly recapitulate the basic principle, let us consider an atom moving with velocity $v$ in one dimension in the presence of two near resonant, counter-propagating laser beams.
 The frequency $\omega$ of these laser beams, which is the same for both beams, is swept symmetrically around the atomic transition frequency $\omega_\mathrm{res}$ in a sawtooth pattern as shown in Fig.~\ref{fig:figure1a}(c). For an atom moving with velocity $v$, the frequencies of the beams will appear Doppler shifted in opposite directions. This induces time ordering in the absorption of the photons from the two counter-propagating beams. For increasing frequency ramps, the beam counter-propagating to the moving atom, will first induce an adiabatic transfer of the atom from the ground-state to the upper state. 
After a specific amount of time, which is determined by the velocity of the atom and the slope of the frequency ramp, the co-propagating beam induces rapid adiabatic passage from the excited state back to the ground-state. During both events, one photon momentum is transferred to the atom in the direction opposite to its motion. The jump in frequency back to red detuning must be diabatic such that the atom stays in the ground-state. In this way, an average force of 
\begin{equation} 
F_\mathrm{SWAP}=2\hbar k f_\mathrm{ramp}
\label{eq: SWAP force}
\end{equation} is acting on the atom, where $k$ is the wavenumber of the used transition and $f_\mathrm{ramp}$ is the repetition rate of the frequency ramp.
If one inverts the slope of the sawtooth ramp, the time ordering of the adiabatic passages is inverted and the atom gets accelerated. To be in the adiabatic regime, the condition 
\begin{equation}
\kappa=\frac{\Omega_{0}^2}{\alpha}=\frac{\Omega_{0}^2}{f_\mathrm{ramp} \Delta_\mathrm{ramp}} \gg 1
\end{equation}
needs to be fulfilled \cite{bartolotta2018laser}. 
\begin{figure}[ht!]
	\centering
		\includegraphics{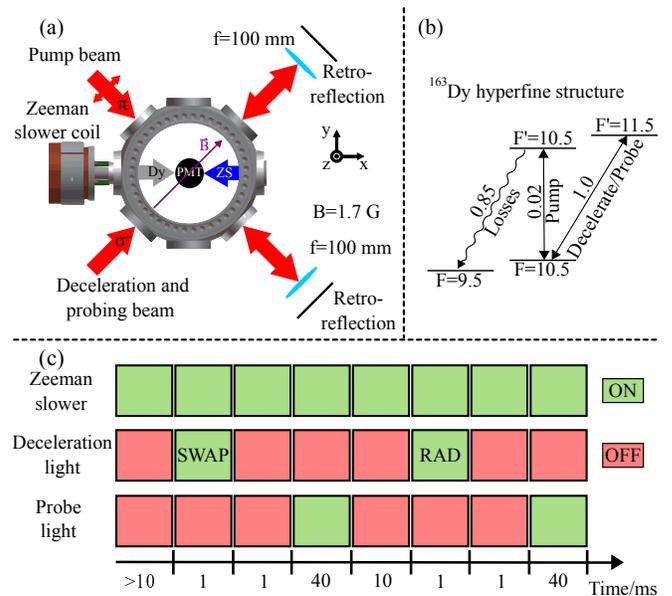}
		\caption{(Color online) (a) Schematic of the vacuum chamber inside which the SWAP/RAD deceleration takes place. The gray arrow labeled ``Dy" indicates the direction of atoms leaving the ZS after deceleration by the ZS beam labeled ``ZS". The direction of the magnetic field used as quantization axis is drawn in purple and the retro-reflected pump and probe beams are indicated by red arrows along with their polarizations. The photomultiplier tube used for the fluorescence measurements is placed below the vacuum chamber and is labeled ``PMT". (b) The part of the $^{163}\mathrm{Dy}$ hyperfine structure that is relevant for the described experimental scheme.
The numbers next to the transition arrows indicate the relative transition strengths normalized to the $\mathrm{F}=10.5 \rightarrow \mathrm{F'}=11.5$ transition. (c) Sequence used to compare the RAD and SWAP force. When losses are added to the $\mathrm{F}=10.5$ ground-state, the pump beam is switched on during the deceleration.}
	\label{fig:figure2}
\end{figure}
Here, $\kappa$ is the adiabaticity parameter, $\alpha$ is the slope of the ramp, $\Delta_\mathrm{ramp}$ is the ramp amplitude and  $\Omega_0$ is the on-resonance Rabi frequency. 

\section{EXPERIMENTAL SCHEME}

The experimental set-up is depicted in Fig.~\ref{fig:figure2}(a). A continuous beam of decelerated Dy atoms leaves a Zeeman slower (ZS) described in \cite{muhlbauer2018systematic}. The atoms have a typical velocity distribution featuring a peak at about $20\,\mathrm{m/s}$ and a tail extending to even negative velocities as shown in Fig.~\ref{fig:figure1app}(b) and (d) in the appendix.
Two circularly polarized laser beams counter-propagate collinearly with a homogeneous $1.7\,\mathrm{G}$ magnetic field directed at $45^{\circ}$ relative to the ZS axis. They are turned on, typically, for $1\, \mathrm{ms}$. 
 The beams are either red detuned from the $626\,\mathrm{nm}$ transition of Dy to act as an optical molasses, or are modulated in the sawtooth manner shown in Fig.~\ref{fig:figure1a}(c) to decelerate the atoms. Their $1/e^2$ beam diameter is $1.7\,\mathrm{cm}$ and the beams pass a $1.4\,\mathrm{cm}$ diameter aperture before entering the vacuum chamber. After deceleration, the laser beams are switched off for $1\,\mathrm{ms}$ as shown in Fig.~\ref{fig:figure2}(c). 
\begin{figure}[ht]
	\centering
		\includegraphics{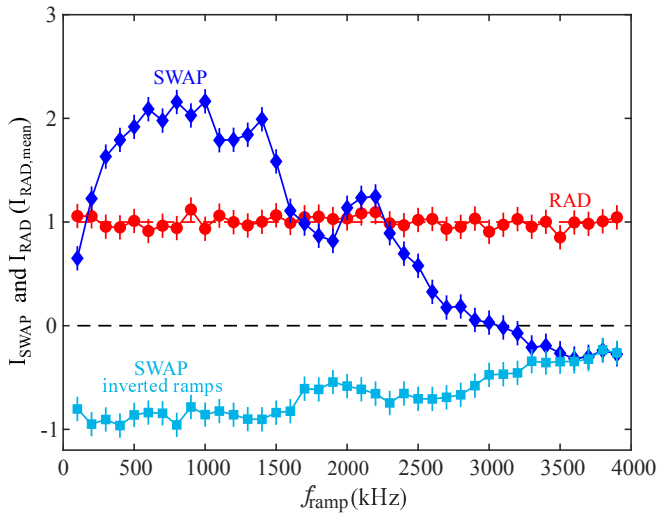}
		\caption{(Color online) SWAP deceleration is applied for $1\,\mathrm{ms}$ for varying ramp repetition rates $f_\mathrm{ramp}$ (dark blue diamonds) and subsequently $\mathrm{I}_\mathrm{SWAP}$ is measured. For comparison, RAD deceleration is applied for the same amount of time and the same saturation parameter with an optimized red detuning $\Delta_\mathrm{RAD}=-15.5 ~\Gamma_{626}$ (red circles). The measured intensities are normalized to the mean intensity after RAD deceleration $I_\mathrm{RAD,mean}$. When the slope of the sawtooth ramp is inverted, we measure negative background-subtracted fluorescence intensities (light blue squares). To estimate the error of the measured intensities we perform $270$ measurements at $f_{\mathrm{ramp}}=400\,\mathrm{kHz}$ and determine the standard deviation $\sigma$ of $\mathrm{I}_\mathrm{SWAP}$ and $\mathrm{I}_\mathrm{RAD}$. The error bars in this figure are the standard error $\sigma/\sqrt{N}$ with $N=15$ being the number of measurements done for each $f_{\mathrm{ramp}}$ setting.}
	\label{fig:figure3}
\end{figure}
They are then switched on again with a typical intensity corresponding to a saturation parameter $\mathrm{S}=\mathrm{I}/\mathrm{I}_\mathrm{S}=24$ and on resonance to the $0\,\mathrm{m/s}$ velocity class for probing the amount of slow atoms. 
The $0\,\mathrm{m/s}$ velocity class atoms exhibit a Doppler-free fluorescence peak in this experimental geometry. This fluorescence is measured with a photomultiplier tube for $40\,\mathrm{ms}$ and the initial intensity, which decays due to atoms moving out of the probe beam in typically $6\,\mathrm{ms}$, is taken as a measure of the amount of atoms at $0\,\mathrm{m/s}$. For further analysis, the background signal, which stems partially from atoms that have been slowed to $0\,\mathrm{m/s}$ by the ZS prior to the RAD/SWAP deceleration, is subtracted.
In the data presented in section~\ref{sec:Exp results}, this background-subtracted fluorescence intensity of the $0\,\mathrm{m/s}$ velocity class is plotted either as absolute values $\mathrm{I}_\mathrm{SWAP}$ or $\mathrm{I}_\mathrm{RAD}$ after SWAP or RAD deceleration, respectively, or as the ratio $\mathrm{I}_\mathrm{SWAP}/\mathrm{I}_\mathrm{RAD}$. 
To compare the SWAP and RAD forces directly, the deceleration and probing sequence is repeated alternatingly for RAD and SWAP. In case of $^{163}\text{Dy}$ the ZS pumps the atoms to the $\mathrm{F}=10.5$ hyperfine ground-state. 
Losses can be added to the ground-state by applying a pump beam on the $\mathrm{F}=10.5 \rightarrow \mathrm{F'}=10.5$ transition during the deceleration as shown in Fig.~\ref{fig:figure2}(b).
More details on the Doppler-free fluorescence signals and the SWAP set-up can be found in the appendix.
\begin{figure}[ht]
	\centering
		\includegraphics{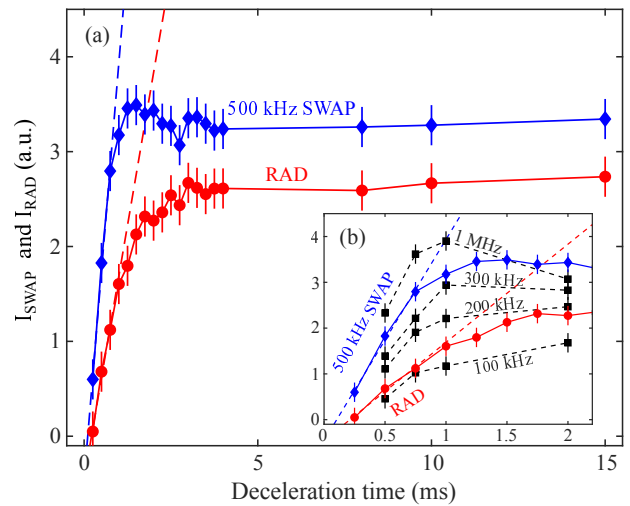}
		\caption{(Color online) (a) $\mathrm{I}_\mathrm{SWAP}$ (dark blue diamonds) and $\mathrm{I}_\mathrm{RAD}$ (red circles) are compared for increasing deceleration time. The SWAP/RAD beam settings and the error calculations are the same as for the data in Fig.~\ref{fig:figure3}. The dashed lines are the result of a linear fit to the first three data points. (b) A zoom into the data points of (a) up to $2\,\mathrm{ms}$. In addition, results for SWAP deceleration with different ramp repetition rates but equal velocity-capture ranges are shown as black squares.}
	\label{fig:figure4}
\end{figure}

\section{EXPERIMENTAL RESULTS}
\label{sec:Exp results}

Unless otherwise noted, all measurements presented here are conducted using $^{163}\text{Dy}$ but similar results were also achieved with $^{162}\text{Dy}$. In Fig.~\ref{fig:figure3} we compare the fluorescence intensities $\mathrm{I}_\mathrm{SWAP}$ and $\mathrm{I}_\mathrm{RAD}$ for different ramp repetition rates $f_\mathrm{ramp}$. Both forces are generated with beams having the same peak saturation parameter $\mathrm{S}=2600$. In the case of the RAD, an optimized red detuning of $\Delta_\mathrm{RAD}=-15.5 ~\Gamma_{626}$ is applied. For SWAP, we use a ramp amplitude of $\Delta_\mathrm{ramp}=190~\Gamma_{626}$ with a zero mean detuning.
Initially, the amount of slow atoms increases with $f_\mathrm{ramp}$ as expected from Eq.~(\ref{eq: SWAP force}).
At $f_\mathrm{ramp}=200\,\mathrm{kHz}$, the SWAP force surpasses the RAD force and at $f_\mathrm{ramp}=500\,\mathrm{kHz}$, it is twice as large. Further increase of $f_\mathrm{ramp}$ does not lead to significantly higher numbers of $0\,\mathrm{m/s}$ atoms and beyond $f_\mathrm{ramp}=1.5\,\mathrm{MHz}$, the amount of slow atoms decreases. This can be related to the decrease of the adiabaticity parameter, which is $\kappa=3.9$ at $1.5\,\mathrm{MHz}$ compared to $\kappa=11.7$ at $500\,\mathrm{kHz}$.
When we invert the slope of the sawtooth ramps, the amount of atoms in the $0\,\mathrm{m/s}$ velocity class decreases to negative values, which means that the atoms are now accelerated as discussed above.
For $f_\mathrm{ramp}\geq 1.5\,\mathrm{MHz}$ less atoms are accelerated in accordance with the less efficient deceleration for rising ramps in this regime. 

To further explore the relative efficiency of SWAP deceleration versus RAD deceleration, we compare the two techniques for different deceleration times at $f_\mathrm{ramp}=500\,\mathrm{kHz}$. The results are shown in Fig.~\ref{fig:figure4}. 
\begin{figure*}[ht]
	\centering
		\includegraphics{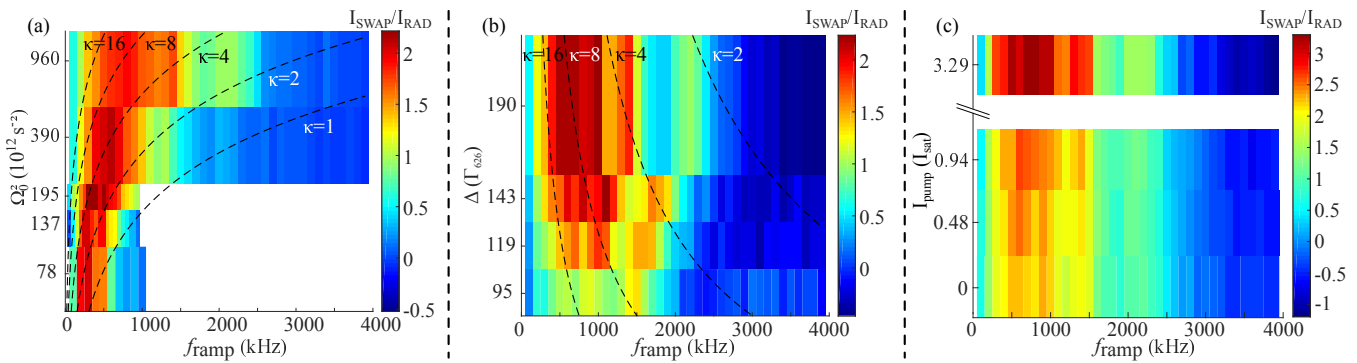}
		\caption{(Color online) All three graphs show the ratio $\mathrm{I}_\mathrm{SWAP}/\mathrm{I}_\mathrm{RAD}$ as a measure of the SWAP deceleration efficiency. In (a) and (b) the dashed lines indicate lines of a constant adiabaticity parameter $\kappa$. (a) For a fixed SWAP ramp amplitude $\Delta_\mathrm{ramp}=190~\Gamma_{626}$ the squared Rabi frequency $\Omega_0^2$ and the ramp repetition rate $f_\mathrm{ramp}$ are varied. For this data set, $^{162}\mathrm{Dy}$ is used. (b) For a fixed $\Omega_0^2=860 \times 10^{12}\mathrm{s}^{-2}$ the ramp amplitude $\Delta_\mathrm{ramp}$ and $f_\mathrm{ramp}$ are varied. (c) The intensity of the pump beam $\mathrm{I}_{\mathrm{pump}}$ which induces losses to the $\mathrm{F}=10.5$ ground-state is increased stepwise and plotted versus $f_\mathrm{ramp}$. The deceleration time is $1\,\mathrm{ms}$ for the data presented in (a) and (c) and $0.75\,\mathrm{ms}$ for the data plotted in (b).}
	\label{fig:figure5}
\end{figure*}
Both forces lead to a linear increase of slow atoms over the first $0.75\,\mathrm{ms}$ until saturation sets in above about $1\,\mathrm{ms}$. The saturation could be due to an equilibrium reached between new atoms entering the deceleration beam region and atoms leaving the beam region due to remaining velocity components perpendicular to the deceleration beams and due to gravity. An increase of slow atoms for deceleration times longer than $6\,\mathrm{ms}$ is not expected since the fluorescence signals decay to the background level on a comparable time scale. The saturated $\mathrm{I}_\mathrm{SWAP}$ is about $24\,\%$ larger than the saturated $\mathrm{I}_\mathrm{RAD}$ which indicates that the SWAP force decelerates atoms from the ZS beam more efficiently. From the slope of the linear range we conclude that the SWAP force decelerates a factor of $2.1$ more atoms per time than the RAD force does. The SWAP force at $500\,\mathrm{kHz}$ is expected to be $2.3$ times stronger than the RAD force for the saturated $626\,\mathrm{nm}$ transition (Eq.~(\ref{eq: RAD force})).
 A larger velocity-capture range would also explain a higher slope, but the capture ranges of the two forces for our set of parameters are theoretically almost equal. In the case of the RAD force the power broadened capture range is $v_\mathrm{c,RAD} =\Gamma_{626} \sqrt{S+1}/k_{626} =4.3\,\mathrm{m/s}$ \cite{metcalf1999laser,metcalf2017colloquium} while in the case of the SWAP force it is $v_\mathrm{c,SWAP}=\Delta_\mathrm{ramp}/(4 k_{626})=4.0\,\mathrm{m/s}$ with $k_{626}$ being the angular wavenumber of the $626\,\mathrm{nm}$ transition \cite{bartolotta2018laser}. 
To check this, we compare SWAP results in Fig.~\ref{fig:figure4}(b) for different ramp repetition rates but equal ramp amplitude $\Delta_\mathrm{ramp}$ and hence equal capture range. The increase of $\mathrm{I}_\mathrm{SWAP}$ for rising ramp repetition rates indicates that the $2.1$ faster accumulation of atoms in the $0\,\mathrm{m/s}$ velocity class is mainly due to a larger force and not due to a larger capture range of the SWAP force. 

In addition to the measurements presented here, we perform an independent optimization of the RAD force for both $^{162}\text{Dy}$ and $^{163}\text{Dy}$ including spectral broadening of the deceleration-beams. With the same deceleration-beam intensity for both forces and a deceleration time of $0.75\,\mathrm{ms}$, no spectral broadening and detuning parameters are found for which the ratio $\mathrm{I}_\mathrm{RAD}/\mathrm{I}_\mathrm{SWAP}$ becomes larger than $0.55$
 with $f_\mathrm{ramp}=500\,\mathrm{kHz}$ and $\Delta_\mathrm{ramp}=190~\Gamma_{626}$.

In Fig.~\ref{fig:figure5} we plot the ratio $\mathrm{I}_\mathrm{SWAP}/\mathrm{I}_\mathrm{RAD}$ for different combinations of parameters ($\Omega_0$, $f_\mathrm{ramp}$ and $\Delta_\mathrm{ramp}$) as a measure of the efficiency of SWAP deceleration over RAD deceleration.
For these measurements, the RAD beams are set to a fixed red detuning of $\Delta_\mathrm{RAD}=-15.5 ~\Gamma_{626}$ and their intensity is equal to the intensity of the SWAP beams. For the data in Fig.~\ref{fig:figure5}(a) we vary the Rabi frequency $\Omega_0$ of the SWAP and RAD beams while $\Delta_\mathrm{ramp}=190~\Gamma_{626}$ is fixed and for Fig.~\ref{fig:figure5}(b) we vary the ramp amplitude $\Delta_\mathrm{ramp}$ while $\Omega_0^2=860 \times 10^{12}\mathrm{s}^{-2}$ ($\mathrm{S}=2350$) is fixed. For reference, lines at which $\kappa$ is constant are plotted also. The data shown in Fig.~\ref{fig:figure5}(a) demonstrate that the SWAP force maintains a factor of two higher deceleration rates than the RAD force over a large range of Rabi frequencies. Additionally, one observes that the mean position of the maximum $\mathrm{I}_{\mathrm{SWAP}}/\mathrm{I}_{\mathrm{RAD}}$ shifts to higher ramp repetition rates when higher $\Omega_0^2$ are applied as expected from the adiabaticity condition, but it does not strictly follow one of the $\kappa=\mathrm{const.}$ lines. For increasing $\Omega_0^2$, the maximum is located at higher $\kappa$ values indicating that not only the adiabaticity condition is limiting the highest usable ramp repetition rates. The data shown in Fig.~\ref{fig:figure5}(b) indicates that even larger $\mathrm{I}_{\mathrm{SWAP}}/\mathrm{I}_{\mathrm{RAD}}$ ratios could be reached for larger $\Delta_\mathrm{ramp}$ values but these are limited in our set-up to about $190~\Gamma_{626}$ for technical reasons. When $\Delta_\mathrm{ramp}$ is increased the $\mathrm{I}_{\mathrm{SWAP}}$ maximum moves to lower $f_\mathrm{ramp}$ values as expected from the adiabaticity condition. The increase of $\mathrm{I}_{\mathrm{SWAP}}$ with larger ramp amplitudes is probably caused by the increase of the capture range of the SWAP force, which is only $2\,\mathrm{m/s}$ at $95~\Gamma_{626}$ compared to $4\,\mathrm{m/s}$ at $190~\Gamma_{626}$. Interestingly, while $\Delta_\mathrm{ramp}$ is increased the absolute $\mathrm{I}_{\mathrm{SWAP}}$ maximum moves to lower $f_\mathrm{ramp}$ values but a second local maximum at higher $f_\mathrm{ramp}$ moves further to higher $f_\mathrm{ramp}$ values in the opposite direction.

The development of new laser cooling techniques, which are robust against population loss from the upper state and lower state of the used transition, is of great interest for experiments with atoms or molecules, which have complex electronic structures with many possible decay channels. 
To simulate the robustness against an unstable ground-state, we actively pump population out of the ground-state into a level not used for cooling. To this end, we add losses to the $\mathrm{F}=10.5$ ground-state of $^{163}\text{Dy}$ during SWAP and RAD deceleration by applying a pump beam that is resonant with the $\mathrm{F}=10.5 \rightarrow \mathrm{F'}=10.5$ transition as shown in Fig.~\ref{fig:figure2}(b). The amount of losses is increased stepwise by increasing the pump beam intensity $\mathrm{I}_{\mathrm{pump}}$ and the results are shown in Fig.~\ref{fig:figure5}(c).
While after both, SWAP and RAD deceleration, losses of slow ground-state atoms are observed, the ratio 
$\mathrm{I}_{\mathrm{SWAP}}/\mathrm{I}_{\mathrm{RAD}}$ increases by $50~\mathrm{\%}$ for $\mathrm{I}_{\mathrm{pump}}=3.29\,\mathrm{I}_{\mathrm{sat}}$ at $f_\mathrm{ramp}=700\,\mathrm{kHz}$ compared to $\mathrm{I}_{\mathrm{pump}}=0$. Thus the SWAP force is less influenced than the RAD force by the added loss mechanism. In order to rule out possible coherent two photon processes from the pump and cooling beams, we add phase noise to the pump laser beam with a $3\,\mathrm{dB}$ bandwidth of $6\,\mathrm{MHz}$ and phase noise amplitudes ranging from $0$ to $2\pi$. Neither $\mathrm{I}_{\mathrm{SWAP}}$ nor $\mathrm{I}_{\mathrm{RAD}}$ is influenced by adding phase noise at $\mathrm{I}_{\mathrm{pump}}=3.29\,\mathrm{I}_{\mathrm{sat}}$ and $f_\mathrm{ramp}=500\,\mathrm{kHz}$.

\section{CONCLUSIONS}

We have demonstrated one dimensional deceleration of Dy by the recently developed SWAP technique. For identical beam intensities and similar velocity-capture ranges, we observe that the SWAP force is by a factor of $2.1$ times faster in decelerating atoms than the RAD force. Furthermore, we observe a higher robustness of the SWAP force against ground-state losses, which were induced by opening a decay channel to another hyperfine ground-state. Our study of the SWAP parameter range should facilitate the integration of SWAP forces into experiments with various other atomic species as well as molecules.

\section{ACKNOWLEDGEMENTS}

The authors would like to thank Carina Baumg\"artner and Lena Maske for their contributions to the experiment.
We gratefully acknowledge financial support by the JGU-Startup funding, DFG-Grossger\"at INST 247/818-1 FUGG and the Graduate School of Excellence MAINZ.

\appendix

\section{Velocity-selective saturated fluorescence spectroscopy}

\begin{figure}[ht!]
	\centering
		\includegraphics{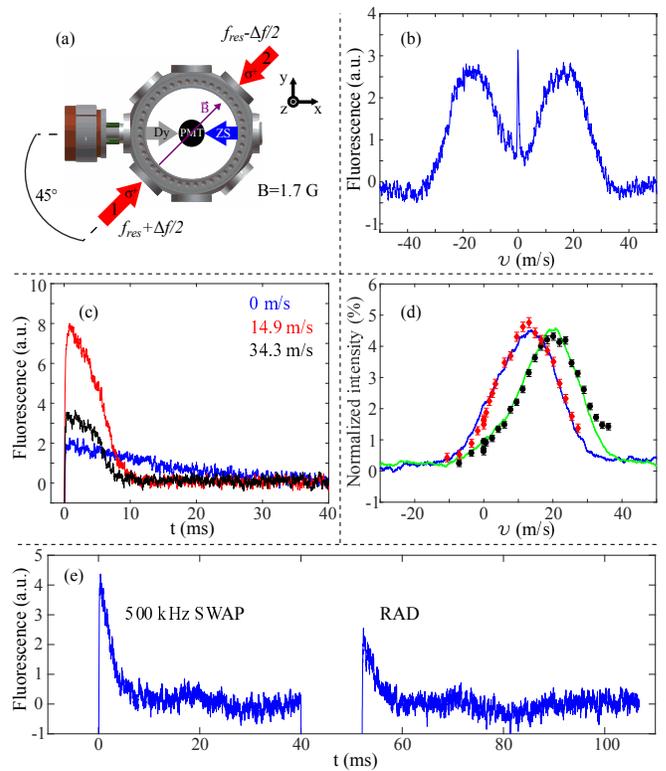}
		\caption{(Color online) (a) Sketch of the set-up used for measuring the velocity distribution of atoms leaving the ZS. The velocity measurement is done in $45^\circ$ to the $x$-axis. (b) Fluorescence spectrum obtained by scanning two counter-propagating laser beams as shown in (a) with $\Delta f=0$. The velocity is given under the assumption that the atoms mainly move into the $x$-direction. The Doppler-free fluorescence signal is visible at $0\,\mathrm{m/s}$ and its FWHM is about $0.45\,\mathrm{m/s}$. (c) Decaying fluorescence signals of atoms with $0\,\mathrm{m/s}$, $14.9\,\mathrm{m/s}$ and $34.3\,\mathrm{m/s}$ velocity in x-direction plotted in blue, red and black, respectively, after suddenly switching off the ZS. The initial intensity is measured by averaging the signals over a time span of $0.6\,\mathrm{ms}$ after switching on the probe beams at time $\mathrm{t}=0$. (d) For two different current settings of the ZS two continuous velocity spectra in blue and green are measured by using and scanning only beam 2 marked in (a). The red and black data points are results for the same ZS settings but are measured by VSSFS. The normalization of these intensities is described in the main text. (e) Typical fluorescence signals measured with the sequence outlined in Fig.~\ref{fig:figure2}(c) to obtain $\mathrm{I}_{\mathrm{SWAP}}$ and $\mathrm{I}_{\mathrm{RAD}}$.} 
	\label{fig:figure1app}
\end{figure}
To determine the number of atoms in the $0\,\mathrm{m/s}$ velocity class, we use a method similar to velocity-selective saturated fluorescence spectroscopy (VSSFS) as used by Gao et al. \cite{gao2014precision}.
As depicted in Fig.~\ref{fig:figure1app}(a), we use two counter-propagating laser beams at $45\,^\circ$ to the atomic beam. When the two laser beams are symmetrically detuned  by $\pm \Delta f/2$ around the atomic resonance frequency $f_{\mathrm{res}}$ of the $626\,\mathrm{nm}$ transition, only atoms with the velocity
\begin{equation}
v=\lambda_{626} \Delta f/[2\, \mathrm{cos}(45^\circ)]
\end{equation}
are resonant with both beams at the same time. Here the atoms are assumed to move on average into the $x$-direction along the ZS axis. When both beams are scanned in frequency while maintaining a fixed relative detuning $\Delta f=0$, one obtains the velocity spectrum shown in Fig.~\ref{fig:figure1app}(b), where two broad peaks are visible. 
The fluorescence of the right (left) peak is caused by laser beam number 1 (2) and the sharp peak in the center is a Doppler-free fluorescence signal with a FWHM of $0.45\,\mathrm{m/s}$ caused by both beams. 

To perform the atom-number measurements presented in the main text the laser frequency is stabilized  to the $0\,\mathrm{m/s}$ velocity class at $\Delta f=0$.
By applying probe beams with varying  $\Delta f$ after switching off the ZS, one can measure the decaying fluorescence of different velocity classes. The initial fluorescence after switching off the ZS is proportional to the amount of atoms in the respective velocity class associated with the chosen $\Delta f$. 
In Fig.~\ref{fig:figure1app}(c) three such fluorescence signals are shown for $0\,\mathrm{m/s}$, $14.9\,\mathrm{m/s}$ and $34.3\,\mathrm{m/s}$. For increasing velocity, the temporal shape of the fluorescence signals becomes more box-like. This is expected for a continuous stream of atoms, which have a certain velocity, that is cut off suddenly. The time of the decay to the background level agrees well with the velocities associated with the used detunings and the distance of the last ZS coil to the end of the probe-beam interaction region, which is about $23\,\mathrm{cm}$. To demonstrate that VSSFS is suitable to selectively measure the amount of atoms in certain velocity classes, we compare standard one-beam Doppler spectroscopy to the results when VSSFS is used in Fig.~\ref{fig:figure1app}(d) and observe a good agreement between the two methods. The VSSFS and the standard Doppler spectra are normalized to the integral over the velocities covered by the VSSFS measurements. Typical fluorescence signals of the $0~\mathrm{m/s}$ velocity class are plotted in Fig.~\ref{fig:figure1app}(e). $\mathrm{I}_{\mathrm{SWAP}}$ and $\mathrm{I}_{\mathrm{RAD}}$ are measured by averaging the intensity over a time span of $0.6\,\mathrm{ms}$ after switching on the probe beams.

\section{Generation and measurement of sawtooth frequency ramps}

\begin{figure}[hb!]
	\centering
		\includegraphics{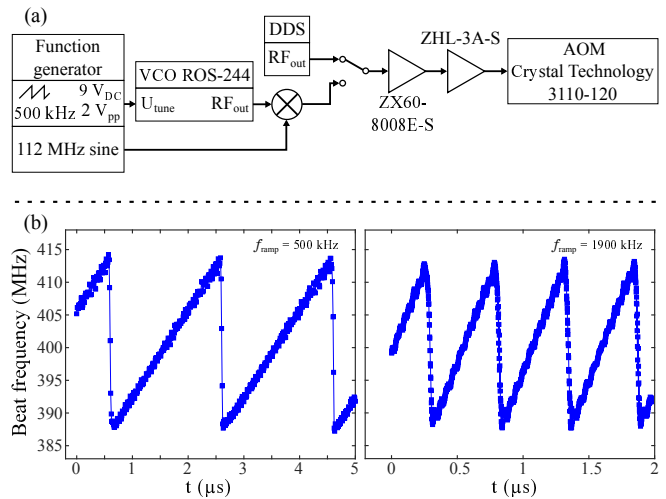}
		\caption{(Color online) (a) Schematic of the RF electronics used to generate the sawtooth frequency ramps at around $95\,\mathrm{MHz}$ to drive the SWAP double-pass AOM. All components except the function generator, AOM and Direct Digital Synthesizer (DDS) are manufactured by Mini-Circuits. (b) Results of an optical beat measurement between a constant-frequency laser beam and a laser beam that is modulated with $\Delta_{\mathrm{ramp}}=190\,\Gamma_{\mathrm{626}}$ and $f_\mathrm{ramp}=500\,\mathrm{kHz}$ (left) and $f_\mathrm{ramp}=1900\,\mathrm{kHz}$ (right) sawtooth waves by using the set-up outlined in (a).} 
	\label{fig:figure2app}
\end{figure}

In this section we describe how a sawtooth frequency modulated laser beam is generated in our set-up and how we switch between the SWAP beam, the RAD beam and the probe beam. To modulate the frequency and control the intensity the laser beam is focused through an Acousto Optic Modulator (AOM) with a $52\,\mu\mathrm{m}$ beam waist in a double-pass configuration.
After passing through the AOM twice, the beam is coupled into a single-mode fiber and guided to the vacuum main chamber. In Fig.~\ref{fig:figure2app}(a) a schematic of the radio-frequency (RF) electronics, which drive the AOM, is shown. There are two sources of AOM driving frequencies.
A Direct Digital Synthesizer generates a $95\,\mathrm{MHz}$ sine wave, which is amplified to drive the AOM during the time when probing is performed.
The second RF source consists of a Voltage Controlled Oscillator (VCO), which is controlled by the output of a function generator. 
The function generator drives the VCO with either sawtooth waves for the SWAP deceleration or with a constant voltage for RAD deceleration.
The VCO output frequency at about $207\,\mathrm{MHz}$ is mixed down to $95\,\mathrm{MHz}$ before it is amplified and sent to the AOM while the last amplifier also acts as a low-pass filter, which cuts off the higher frequencies leaving the mixer. With the RF switch we choose between the probe beam and the SWAP/RAD beam setting. 
To verify that this set-up can modulate the laser beam with sawtooth frequency ramps, we measure the beat between a constant-frequency laser beam and the SWAP beam on a photodiode. The frequency of the beat signal is for technical reasons centered around $400\,\mathrm{MHz}$. By fitting sine functions to $25\,\mathrm{ns}$ long segments of the beat signal, the instantaneous beat frequency is obtained. The results for two settings of the ramp repetition rate are plotted in Fig.~\ref{fig:figure2app}(b) versus time and show the expected sawtooth form. 

\bibliography{SWAP_bib}{}

\begin{thebibliography}{26}
\providecommand{\natexlab}[1]{#1}
\providecommand{\url}[1]{\texttt{#1}}
\expandafter\ifx\csname urlstyle\endcsname\relax
  \providecommand{\doi}[1]{doi: #1}\else
  \providecommand{\doi}{doi: \begingroup \urlstyle{rm}\Url}\fi

\bibitem[Metcalf and Van~der Straten(1999)]{metcalf1999laser}
H.~Metcalf and P.~Van~der Straten.
\newblock Laser cooling and trapping.
\newblock \emph{Springer, New York}, 1999.

\bibitem[Ketterle et~al.(1999)Ketterle, Durfee, and
  Stamper-Kurn]{ketterle1999making}
W.~Ketterle, D.~S. Durfee, and D.~M. Stamper-Kurn.
\newblock Making, probing and understanding {B}ose-{E}instein condensates.
\newblock \emph{arXiv:cond-mat/9904034}, 1999.

\bibitem[Ketterle and Zwierlein(2008)]{ketterle2008making}
W.~Ketterle and M.~W. Zwierlein.
\newblock Making, probing and understanding ultracold {F}ermi gases.
\newblock \emph{arXiv:0801.2500}, 2008.

\bibitem[Eschner et~al.(2003)Eschner, Morigi, Schmidt-Kaler, and
  Blatt]{eschner2003laser}
J.~Eschner, G.~Morigi, F.~Schmidt-Kaler, and R.~Blatt.
\newblock Laser cooling of trapped ions.
\newblock \emph{Journal of the Optical Society of America B}, 20\penalty0
  (5):\penalty0 1003--1015, 2003.

\bibitem[Ludlow et~al.(2015)Ludlow, Boyd, Ye, Peik, and
  Schmidt]{ludlow2015optical}
A.~D. Ludlow, M.~M. Boyd, J.~Ye, E.~Peik, and P.~O. Schmidt.
\newblock Optical atomic clocks.
\newblock \emph{Reviews of Modern Physics}, 87\penalty0 (2):\penalty0 637,
  2015.

\bibitem[Metcalf(2017)]{metcalf2017colloquium}
H.~Metcalf.
\newblock Colloquium: {S}trong optical forces on atoms in multifrequency light.
\newblock \emph{Reviews of Modern Physics}, 89\penalty0 (4):\penalty0 041001,
  2017.

\bibitem[Voitsekhovich et~al.(1989)Voitsekhovich, Danileiko, Negriiko,
  Romanenko, and Yatsenko]{voitsekhovich1989observation}
V.~S. Voitsekhovich, M.~V. Danileiko, A.~Negriiko, V.~Romanenko, and
  L.~Yatsenko.
\newblock Observation of a stimulated radiation pressure of amplitude-modulated
  light on atoms.
\newblock \emph{JETP Letters}, 49:\penalty0 161--164, 1989.

\bibitem[S{\"o}ding et~al.(1997)S{\"o}ding, Grimm, Ovchinnikov, Bouyer, and
  Salomon]{soding1997short}
J.~S{\"o}ding, R.~Grimm, Y.~B. Ovchinnikov, P.~Bouyer, and C.~Salomon.
\newblock Short-distance atomic beam deceleration with a stimulated light
  force.
\newblock \emph{Physical Review Letters}, 78\penalty0 (8):\penalty0 1420, 1997.

\bibitem[Partlow et~al.(2004)Partlow, Miao, Bochmann, Cashen, and
  Metcalf]{partlow2004bichromatic}
M.~Partlow, X.~Miao, J.~Bochmann, M.~Cashen, and H.~Metcalf.
\newblock Bichromatic slowing and collimation to make an intense helium beam.
\newblock \emph{Physical Review Letters}, 93\penalty0 (21):\penalty0 213004,
  2004.

\bibitem[Yatsenko and Metcalf(2004)]{yatsenko2004dressed}
L.~Yatsenko and H.~Metcalf.
\newblock Dressed-atom description of the bichromatic force.
\newblock \emph{Physical Review A}, 70\penalty0 (6):\penalty0 063402, 2004.

\bibitem[Corder et~al.(2015)Corder, Arnold, and Metcalf]{corder2015laser}
C.~Corder, B.~Arnold, and H.~Metcalf.
\newblock Laser cooling without spontaneous emission.
\newblock \emph{Physical Review Letters}, 114\penalty0 (4):\penalty0 043002,
  2015.

\bibitem[Lu et~al.(2005)Lu, Miao, and Metcalf]{lu2005bloch}
T.~Lu, X.~Miao, and H.~Metcalf.
\newblock Bloch theorem on the {B}loch sphere.
\newblock \emph{Physical Review A}, 71\penalty0 (6):\penalty0 061405, 2005.

\bibitem[Miao et~al.(2007)Miao, Wertz, Cohen, and Metcalf]{miao2007strong}
X.~Miao, E.~Wertz, M.~G. Cohen, and H.~Metcalf.
\newblock Strong optical forces from adiabatic rapid passage.
\newblock \emph{Physical Review A}, 75\penalty0 (1):\penalty0 011402, 2007.

\bibitem[Stack et~al.(2011)Stack, Elgin, Anisimov, and
  Metcalf]{stack2011numerical}
D.~Stack, J.~Elgin, P.~M. Anisimov, and H.~Metcalf.
\newblock Numerical studies of optical forces from adiabatic rapid passage.
\newblock \emph{Physical Review A}, 84\penalty0 (1):\penalty0 013420, 2011.

\bibitem[Norcia et~al.(2018)Norcia, Cline, Bartolotta, Holland, and
  Thompson]{norcia2018narrow}
M.~A. Norcia, J.~R.~K. Cline, J.~P. Bartolotta, M.~J. Holland, and J.~K.
  Thompson.
\newblock Narrow-line laser cooling by adiabatic transfer.
\newblock \emph{New Journal of Physics}, 20\penalty0 (2):\penalty0 023021,
  2018.

\bibitem[Muniz et~al.(2018)Muniz, Norcia, Cline, and Thompson]{muniz2018robust}
J.~A. Muniz, M.~A. Norcia, J.~R.~K. Cline, and J.~K. Thompson.
\newblock A {R}obust {N}arrow-{L}ine {M}agneto-{O}ptical {T}rap using
  {A}diabatic {T}ransfer.
\newblock \emph{arXiv:1806.00838}, 2018.

\bibitem[Greve et~al.(2018)Greve, Wu, and Thompson]{greve2018laser}
G.~P. Greve, B.~Wu, and J.~K. Thompson.
\newblock Laser {C}ooling with {A}diabatic {T}ransfer on a {R}aman
  {T}ransition.
\newblock \emph{arXiv:1805.04452}, 2018.

\bibitem[Martin et~al.(1978)Martin, Zalubas, and Hagan]{martin1978}
W.~C. Martin, R.~Zalubas, and L.~Hagan.
\newblock Atomic energy levels-{T}he rare-{E}arth elements.
\newblock \emph{NSRDS-NBS, Washington: National Bureau of Standards, US
  Department of Commerce,| c1978}, 1978.

\bibitem[Hogervorst et~al.(1978)Hogervorst, Zaal, Bouma, and
  Blok]{hogervorst1978isotope}
W.~Hogervorst, G.~J. Zaal, J.~Bouma, and J.~Blok.
\newblock Isotope shifts and hyperfine structure of natural dysprosium.
\newblock \emph{Physics Letters A}, 65\penalty0 (3):\penalty0 220--222, 1978.

\bibitem[Gustavsson et~al.(1979)Gustavsson, Lundberg, Nilsson, and
  Svanberg]{gustavsson1979lifetime}
M.~Gustavsson, H.~Lundberg, L.~Nilsson, and S.~Svanberg.
\newblock Lifetime measurements for excited states of rare-earth atoms using
  pulse modulation of a cw dye-laser beam.
\newblock \emph{Journal of the Optical Society of America}, 69\penalty0
  (7):\penalty0 984--992, 1979.

\bibitem[Collopy et~al.(2015)Collopy, Hummon, Yeo, Yan, and
  Ye]{collopy2015prospects}
A.~L. Collopy, M.~T. Hummon, M.~Yeo, B.~Yan, and J.~Ye.
\newblock Prospects for a narrow line {MOT} in {YO}.
\newblock \emph{New Journal of Physics}, 17\penalty0 (5):\penalty0 055008,
  2015.

\bibitem[Kramida et~al.(2018)Kramida, {Yu.~Ralchenko}, Reader, and {NIST ASD
  Team}]{NIST_ASD}
A.~Kramida, {Yu.~Ralchenko}, J.~Reader, and {NIST ASD Team}.
\newblock {NIST Atomic Spectra Database (ver. 5.5.6), [Online]. Available:
  {\tt{https://physics.nist.gov/asd}} [2018, September 11]. National Institute
  of Standards and Technology, Gaithersburg, MD.}, 2018.

\bibitem[Youn et~al.(2010)Youn, Lu, Ray, and Lev]{youn2010dysprosium}
S.~H. Youn, M.~Lu, U.~Ray, and B.~L. Lev.
\newblock Dysprosium magneto-optical traps.
\newblock \emph{Physical Review A}, 82\penalty0 (4):\penalty0 043425, 2010.

\bibitem[Bartolotta et~al.(2018)Bartolotta, Norcia, Cline, Thompson, and
  Holland]{bartolotta2018laser}
J.~P. Bartolotta, M.~A. Norcia, J.~R.~K. Cline, J.~K. Thompson, and M.~J.
  Holland.
\newblock Laser {C}ooling by {S}awtooth {W}ave {A}diabatic {P}assage.
\newblock \emph{arXiv:1806.02931}, 2018.

\bibitem[M{\"u}hlbauer et~al.(2018)M{\"u}hlbauer, Petersen, Baumg{\"a}rtner,
  Maske, and Windpassinger]{muhlbauer2018systematic}
F.~M{\"u}hlbauer, N.~Petersen, C.~Baumg{\"a}rtner, L.~Maske, and
  P.~Windpassinger.
\newblock Systematic optimization of laser cooling of dysprosium.
\newblock \emph{Applied Physics B}, 124\penalty0 (6):\penalty0 120, 2018.

\bibitem[Gao et~al.(2014)Gao, Liu, Xu, Tian, Wang, Ren, Wu, and
  Chang]{gao2014precision}
F.~Gao, H.~Liu, P.~Xu, X.~Tian, Y.~Wang, J.~Ren, H.~Wu, and H.~Chang.
\newblock Precision measurement of transverse velocity distribution of a
  strontium atomic beam.
\newblock \emph{AIP advances}, 4\penalty0 (2):\penalty0 027118, 2014.

\end{thebibliography}
\bibliographystyle{unsrtnat}

\end{document}